\documentclass[10pt]{article}
\usepackage{url}
\usepackage{amsmath}
\usepackage{amsfonts}
\usepackage{natbib}
\usepackage{authblk}
\usepackage[textfont = {footnotesize} , labelfont = bf]{caption}
\usepackage{subfig}
\usepackage{graphicx}
\def \iPAC{\emph{iPAC}}

\begin{document}

\title{Utilizing Protein Structure to Identify Non-Random Somatic Mutations}

\author[1]{Gregory Ryslik}
\author[2]{Yuwei Cheng}
\author[2,3]{Kei-Hoi Cheung}
\author[4]{Yorgo Modis}
\author[1]{Hongyu Zhao}
\affil[1]{Department of Biostatistics, Yale School of Public Health, New Haven, CT, USA}
\affil[2]{Program of Computational Biology and Bioinformatics, Yale University, New Haven, CT, USA}
\affil[3]{Yale Center for Medical Informatics, Yale School of Medicine, New Haven, CT, USA}
\affil[4]{Department of Molecular Biophysics \& Biochemistry, Yale University, New Haven, CT,USA}

\maketitle

\begin{abstract}
\noindent \textbf{Motivation:}
Human cancer is caused by the accumulation of somatic mutations in tumor suppressors and oncogenes within the genome. In the case of oncogenes, recent theory suggests that there are only a few key ``driver" mutations responsible for tumorigenesis. As there have been significant pharmacological successes in developing drugs that treat cancers that carry these driver mutations, several methods that rely on mutational clustering have been developed to identify them. However, these methods consider proteins as a single strand without taking their spatial structures into account. We propose a new methodology that incorporates protein tertiary structure in order to increase our power when identifying mutation clustering.\\
\textbf{Results:}
We have developed a novel algorithm, \iPAC{} (\textbf{i}dentification of \textbf{P}rotein \textbf{A}mino acid \textbf{C}lustering), for the identification of non-random somatic mutations in proteins that takes into account the three dimensional protein structure. By using the tertiary information, we are able to detect both novel clusters in proteins that are known to exhibit mutation clustering as well as identify clusters in proteins without evidence of clustering based on existing methods. For example, by combining the data in the Protein Data Bank (PDB) and the Catalogue of Somatic Mutations in Cancer, our algorithm identifies new mutational clusters in well known cancer proteins such as KRAS and PI3KC$\alpha$. Further, by utilizing the tertiary structure, our algorithm also identifies clusters in EGFR, EIF2AK2, and other proteins that are not identified by current methodology. \\\\
\textbf{Availability:}
R package available on Bioconductor at\\ \url{http://www.bioconductor.org/packages/2.12/bioc/html/iPAC.html}. \\ \\
\textbf{Contacts:} gregory.ryslik@yale.edu; hongyu.zhao@yale.edu 
\end{abstract}

\section{Introduction}
Cancer is one of the most widespread and heterogeneous diseases imposing a huge toll on patients, relatives, friends, and society. However, at its most basic, it is a genetic disease that is caused  by the accumulation of somatic mutations in oncogenes and tumor suppressors \citep{vogelstein_cancer_2004}. While mutations in tumor suppressors tend to down-regulate the activity of genes that prevent cancer, mutations in proto-oncongenes either up-regulate or deregulate the activities of the resulting proteins. So far, pharmacological intervention has shown to be more successful inhibiting the activating oncogenes than restoring tumor suppressing gene function. Coupled with the idea of ``oncogene addiction", that many cancers rely on mutations in a small subset of key genes to be able to continue their uncontrolled growth while the remainder of the mutations constitute passenger mutations \citep{greenman_patterns_2007, weinstein_mechanisms_2006}, the problem of identifying activating oncogenic mutations has received great attention in cancer research.

Recently, several studies have shown support for the hypothesis that activating somatic mutations tend to cluster in protein kinases \citep{torkamani_prediction_2008,greenman_patterns_2007, bardelli_mutational_2003}. Further, as observed by \citet{ye_2010}, mutational clusters might provide further information regarding where to look for activating mutations, reducing the driver mutation search space needed to be analyzed. Moreover, mutational clusters that lead to either beneficial or detrimental phenotypic changes may point to regions that are under positive or directional selection as well as regions that are functionally significant and thus can be targeted by protein engineering \citep{wagner_rapid_2007}.

So far, several methods based upon the number of mutations in a specific region have been developed to detect potential driver oncogenic mutations as well as naturally selected regions. One common method hypothesizes that driver mutations have a higher non-synonymous mutation rate as compared to the background mutation rate \citep{sjoblom_consensus_2006, bardelli_mutational_2003}. Further, one can look at 
the ratio of nonsynonymous ($K_a$) to synonymous ($K_s$) changes per site, $\frac{K_a}{K_s}$ \citep{kreitman_methods_2000}. A criterion for selection is then to check if $\frac{K_a}{K_s} > 1$, based on the hypothesis that the benchmark neutral rate of nucleotide substitution is exceeded when positive selection also contributes to the substitution process. Similarly, \citet{wang_prevalence_2002} proposes a hypothesis that driver mutations have a larger mutational rate than the background mutational rate after gene length normalization. 

While the approaches mentioned above have had some success in detecting positive selection and/or identifying driver mutations, they nevertheless have several shortcomings. First, many of them are dependent on calculating the disparity in non-synonymous versus synonymous mutations but do not recognize that selection often occurs on very small sections of the gene and thus might fail when averaged over the entirety of the gene length. Second, the methods described above \citep{wang_prevalence_2002, kreitman_methods_2000} do not make any attempt to distinguish between activating and non-activating non-synonymous mutations. 

In addition to the approaches described above, some researchers have focused on creating classifiers in order to determine mutation status. As described in \citet{reva_predicting_2011}, these algorithms employ a variety of machine learning techniques, such as Random Forests \citep{breiman_randomforest} and Support Vector Machines \citep{cortes_support-vector_1995}, to calculate a score for each mutation. These scores are typically calculated using a combination of  physico-chemical properties such as evolutionary conservation, size and polarity of substituted and original residues as well as surface accessibility. These scores are then used to classify the mutation. For example, \emph{PolyPhen-2} \citep{adzhubei_method_2010} predicts whether a missense mutation is damaging while \emph{CHASM} \citep{carter_cancer-specific_2009} attempts to discriminate between driver and passenger mutations. While several of these models have had significant success in classifying the mutation, they all require large and well annotated data sets in order to first train the machine learning classifier and then apply the resulting rule set.

Recently, \citet{ye_2010} developed Non-Random Mutational Clustering (NMC) to identify potential activating mutations by hypothesizing that, in the absence of heretofore known mutational hotspots, a mutational cluster is indicative of selection for an activating driver mutation since only a small number of precise mutations can activate a protein \citep{torkamani_prediction_2008, bardelli_mutational_2003}. By looking at the order statistics and assuming that the locations of amino acid mutations follow a uniform distribution when the protein is considered in linear form under the null hypothesis, they identify clusters by calculating whether any two pair-wise mutations are closer together on the line than expected by chance alone.  Despite its success, one limitation of the NMC method is that the proteins are treated as a linear sequence without considering the three dimensional structures of the proteins. 

In this work, we extend the NMC methodology to account for tertiary protein structure. This enables the identification of mutational clusters that are relatively far away in linear space but relatively close together in 3D space.  We proceed to show that our methodology is effective in identifying novel mutational clusters that are missed by NMC in key cancer proteins such as KRAS and PIK3C$\alpha$. Unlike NMC, \iPAC{} is also able to identify the EGFR and EIF2AK2 proteins as containing mutational clustering as well. We also show that many of the clusters identified by \iPAC{} are predicted to be deleterious by well known machine learning algorithms such as Polyphen-2 \citep{adzhubei_method_2010}. However, \iPAC{} has the distinct advantage of requiring only the mutational positions and tertiary structure which allows its application to novel mutations and structures for which extensive information and literature is not yet available. Finally,  we also show that for a large percentage of protein structures, the tertiary structure leads to a net reduction in mutational clusters found, thus presenting a simplified clustering mutational landscape. Ultimately, by providing a refined picture of the mutational clustering, we are are able to provide a more accurate representation of where potential activating mutations may reside within the protein.

\section{Methods}

Our method, named \iPAC{}, uses a 4 step approach to finding mutational clusters. First, mutational and positional data are obtained from the COSMIC \citep{forbes_catalogue_2008} and PDB \citep{pdb} databases (described in Sections \ref{data:COSMIC} and \ref{data:PDB}, respectively). The mutational and positional information is then reconciled to allow a single numerical reference to identify the same physical amino acid in both databases (Section \ref{data:Reconciliation}). Next, MultiDimensional Scaling (MDS) \citep{borg_modern_1997} is used to map the protein structure from 3D to 1D space while preserving, as best as possible, all pairwise three dimensional distances between amino acids for a given protein (Section \ref{Calc:MDS}). The NMC algorithm is then run on the remapped amino acids to find mutational clusters (Section \ref{Calc:NMC}). Finally, the clusters are mapped back into the original protein space and reported back to the user. In the following subsections we discuss each of these steps in detail.

\subsection{Obtaining Mutational Data} \label{data:COSMIC}

Mutational data were obtained from the COSMIC database (version 58) via \url{ftp://ftp.sanger.ac.uk/pub/CGP/cosmic} and implemented using Oracle. In order to  justify the assumption that amino acids follow a uniform distribution of mutation, only mutations that were found through whole gene screens were included. Further, we only used missense mutations that belonged to two categories: 1) ``Confirmed somatic variant" or 2) ``Reported in another cancer sample as somatic". All nonsense and synonymous mutations as well as mutations that had different somatic status categories were excluded. Further, as multiple studies can report mutational data from the same cell line, mutational redundancies were removed to avoid double counting. See ``COSMIC query" in the supplementary information for the SQL code and schema used to generate the data. Finally, in order to  match mutational data with structural data, only the proteins for which a UniProt Accession Number \citep{the_uniprot_consortium_reorganizing_2011} was available were kept. This resulted in 777 unique proteins. 

\subsection{Obtaining the 3D Structural Data} \label{data:PDB}
The protein structural data were obtained from the PDB database via \url{http://www.pdb.org}. As one protein can have several structures, for each of the 777 proteins described above, all the structures with a matching UniProt Accession Number were obtained. If a specific structure had more than one polypeptide chain with a matching amino acid sequence in UniProt, the first matching chain listed was used (typically chain A). For proteins where the resolution was sufficiently high enough to provide more than one alternative conformation for a specific amino acid side chain, only the first conformation listed in the file was used. Once the appropriate side chain and conformation was selected, the $(x,y,z)$ coordinates of all the $\alpha$-carbon atoms were extracted and used to represent the 3D backbone structure of the protein. In all, this process resulted in 1,904 structures. See ``Structure Files" in the supplementary information for a full listing of the structures and side chains used for each protein considered.

\subsection{Reconciling the Structural and Mutational Data} \label{data:Reconciliation}
Due to a different numbering system of the amino acids employed by the PDB and COSMIC databases, an alignment needed to be performed in order to reference the same residue numerically in both databases. Two methods in the \iPAC{} package were designed to reconcile these differences, one based on pairwise alignment \citep{Biostrings_2012} and the other based on a numerical reconstruction from the structural data obtained from the PDB. As there are often significant technical difficulties for such a reconstruction, for the rest of this paper, unless specifically noted, pairwise alignment was used to reconcile these elements. Please see the documentation in the \iPAC{} package for a full description of these two methods. Successful alignment of mutational and positional data occurred on 140 proteins which corresponded to 1100 unique structure/side-chain combinations and 667 unique residue positions containing 1,434 total mutations. We note that for any given structure/side-chain combination, if there is no positional data for a specific residue, the mutational data for that residue is not used. Please see ``Structure Files" in the supplementary information for a full description.

\subsection{Multidimensional Scaling} \label{Calc:MDS}
As the underlying clustering algorithm is dependent upon the construction of order statistics, we used MDS  \citep{borg_modern_1997} to remap the amino acids into one dimensional space while preserving (as best as possible) the pairwise distances between them in 3D space. Specifically, given an $n \times n$ dissimilarity matrix, \[\Delta_{n,n}	 =  \left( \begin{array}{cccc}
\delta_{1,1} & \delta_{1,2} & \cdots &\delta_{1,n} \\
\delta_{2,1} & \delta_{2,2} & \cdots &\delta_{2,n} \\
\vdots & \vdots & \ddots &\vdots \\
\delta_{n,1} & \delta_{n,2} & \cdots & \delta_{n,n} \end{array} \right)\] 
the MDS algorithm maps each $\delta_{i,j}$ into a corresponding distance $d_{i,j}(\mathbf{X})$ on a new m-dimensional metric space $\mathbf{X}$. Formally, for a specific representation function, $f:\delta_{i,j} \rightarrow d_{i,j}(\mathbf{X})$, we have that the original dissimilarities are preserved in $\mathbf{X}$, specifically, $f(\delta_{i,j}) = d_{i,j}(\mathbf{X})$. Here, $f$ can be either fully defined or chosen from a specified class of functions and is employed to handle the case when the proximity measures come from a space that is not necessarily a true metric space. Further, as it is not always possible to preserve the exact distance (for example, due to sampling effects, measurement precision or loss of dimensionality), rather than insist on $f(\delta_{i,j}) = d_{i,j}(\mathbf{X})$, the MDS framework is typically set up such that $f(\delta_{i,j}) \approx d_{i,j}(\mathbf{X})$. Thus, by minimizing a badness-of-fit measure called \emph{raw stress} = $\sigma_r =\sum_{i,j}[f(\delta_{i,j}) - d_{i,j}(\mathbf{x})]^2$, we identify the $\mathbf{x_1,...,x_n}$, that preserve our distances in the new metric space X. However, raw Stress by itself is not always informative as it is subject to distortion by the choice of units used. For instance, if the scale used to measure changes by a factor of 100, the raw stress will change as well but by a factor of $100^2$. Thus, \emph{Stress-1}, which is defined as: 
\begin{eqnarray}
\sigma_1 = \sqrt{\frac{\sum_{i,j}[f(\delta_{i,j}) - d_{i,j}(\mathbf{X})]^2}{\sum_{i,j}d_{i,j}^2(\mathbf{X})}}
\end{eqnarray}
and is not subject to unit distortion, will be minimized instead.

For the purposes of this paper, the dissimilarity matrix is simply equal to pair-wise distance between any two amino acids in the protein. Specifically, the distance between residues $i$ and $j$, denoted $\delta_{i,j}$, is taken to be the Euclidean distance between their respective $\alpha$-carbon atoms. As Euclidean space is a proper metric space, from now on we assume that $f$ is the identity function.  Further, as we require units along the line in order to calculate order statistics, the MDS algorithm will be applied such that we find $\mathbf{x_1}, ...,\mathbf{x_n} \in \mathbb{R}^1$.  Thus, the MDS algorithm finds scalars $x_1,...,x_n$ such that $|x_i - x_j| \approx \delta_{i,j},$ for any two pairwise amino acids $i$ and $j$ in the protein.  We present an example when MDS is applied to the 3GFT structure of KRAS \citep{kras_3GFT} in Figures \ref{fig:KRASRaw} and \ref{fig:KRASMDS} below.

\begin{figure}[htb*]
\begin{minipage}[b]{.5\linewidth}
\centering
\includegraphics[width=\textwidth]{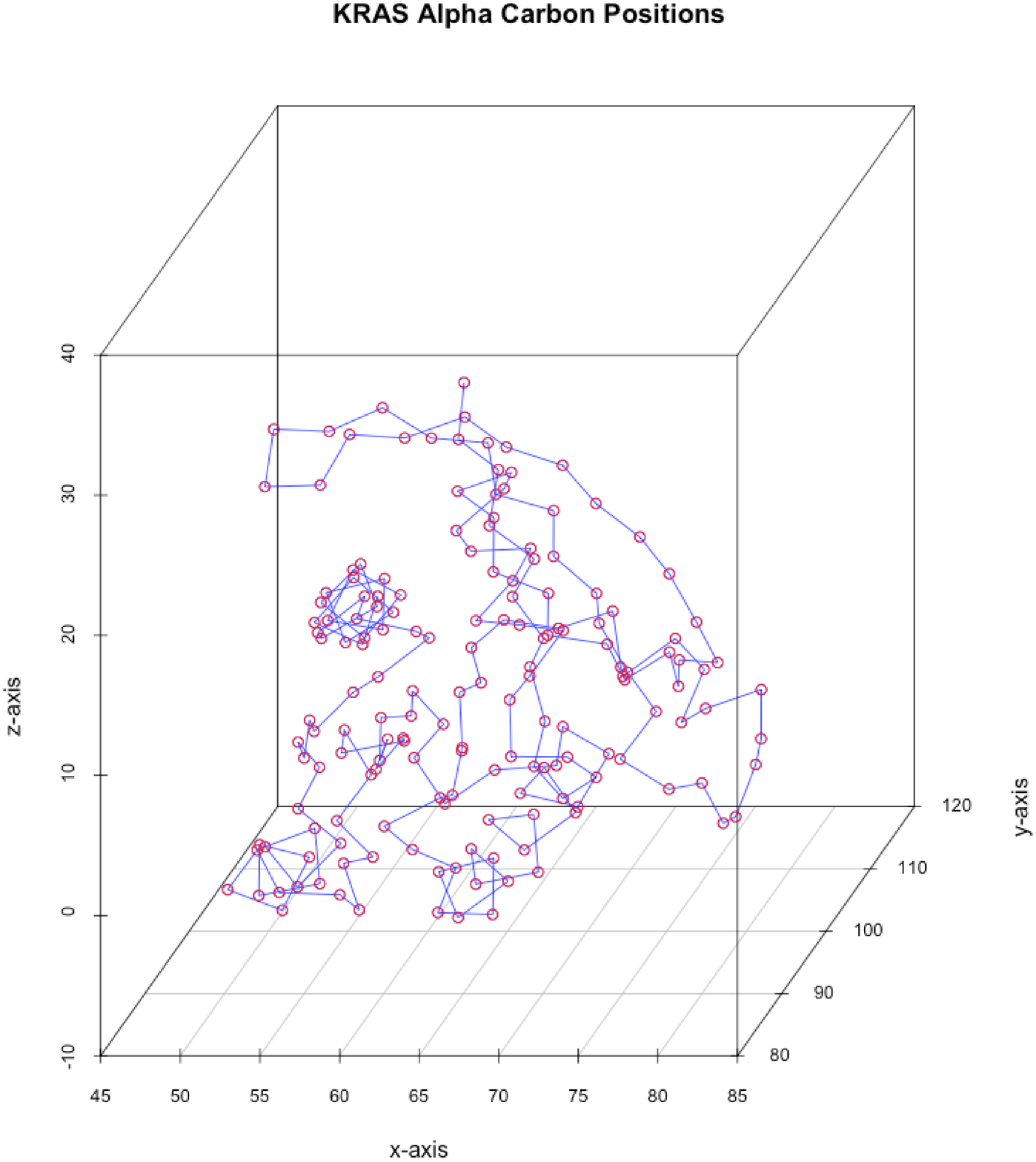}
\caption{KRAS $\alpha$-carbons in 3D Space.}
\label{fig:KRASRaw}
\end{minipage}
\hspace{0.1cm}
\begin{minipage}[b]{0.45\linewidth}
\centering
\includegraphics[width=\textwidth]{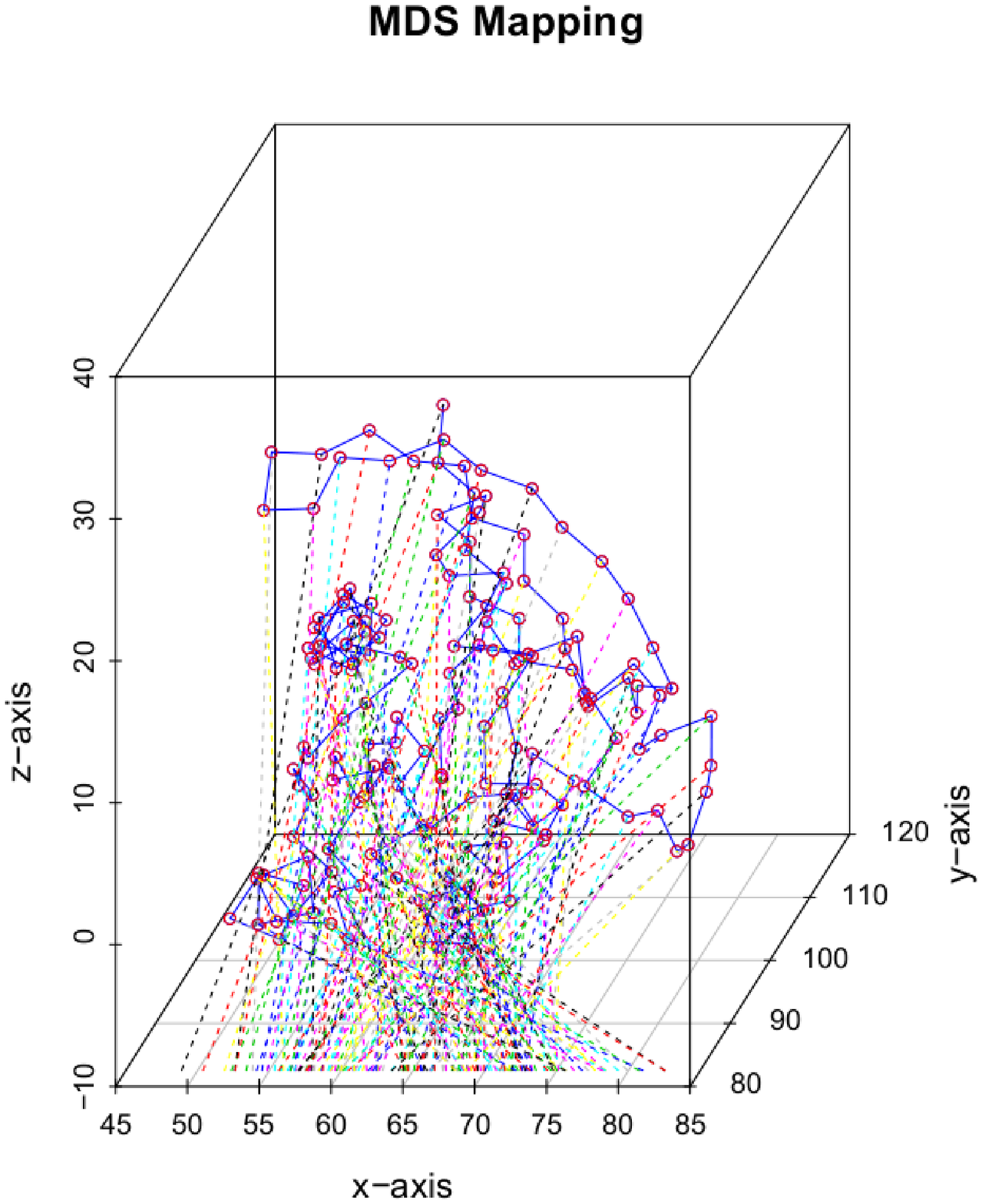}
\caption{KRAS $\alpha$-carbons mapped to the x-axis using MDS.}
\label{fig:KRASMDS}
\end{minipage}
\end{figure} 
 
\subsection{NMC} \label{Calc:NMC}
\begin{figure}[h!]
	\centering
	\includegraphics[scale=0.32]{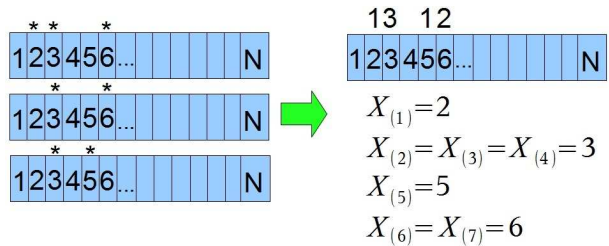}
	\caption{An example of constructing the order statistics. Suppose we had 3 samples of a protein that is N amino acids long. If amino acid $i$ has a ``*" above it, that indicates that the amino acid for that sample had a non-synonymous missense mutation. The samples are then collapsed together and the number of mutations for each residue is shown above the box on the right. These counts form the order statistics. The first mutation is on residue 2 ($X_{(1)} = 2$), the next 3 mutations are on residue 3 ($X_{(2)}=X_{(3)}=X_{(4)} = 3$) , the next mutation is on residue 5 ($X_{(5)} = 5$) and the last 2 mutations are on residue 6 ($X_{(6)}=X_{(7)} = 6$).}
	\label{fig:OrderStats}
\end{figure}

We employed the NMC algorithm \citep{ye_2010} to find the mutational clusters in one dimensional space. Specifically, consider a protein with $N$ amino acids and that each amino acid has a uniform probability of $\frac{1}{N}$ of mutation. Given $m$ samples and $n$ mutations, we are able to calculate the order statistics for every mutation (see Figure \ref{fig:OrderStats}). Two mutations $X_{(i)}$ and $X_{(k)}$ are then defined to be clustered if, $Pr(C_{ki} = X_{(k)}-X_{(i)}) \leq \alpha$. This probability is then calculated for every pair of mutations and adjusted for multiple comparisons using either the Benjamini-Hochberg (BH) adjustment \citep{BH1995} or the Bonferroni adjustment. For the analyses performed in this paper, the more conservative Bonferroni adjustment was used. Finally, it is important to note that the structural information obtained for each protein often does not include positional information on every amino acid within the protein. We removed these ``missing" amino acids from the protein before running the NMC clustering algorithm so that we can compare \iPAC{} and NMC on an equal basis. 

\citet{ye_2010} derive closed form solutions to calculate the $Pr(C_{ki} = c)$ for $c \in \{0, 1,...,N-1\}$. However, as this becomes computationally inefficient, they suggest dividing $C_{ki}$ by N and assuming a continuous uniform distribution on $(0,1)$.  They then show that in the limit, the CDF becomes as follows:

\begin{equation} \label{eqn:continuous}
\begin{aligned}
&Pr(\frac{C_{ki}}{N}=\frac{X_{(k)} - X_{(i)}}{N} \leq c) \\
&= \int_0^c \frac{n!}{(k-i-1)!(i+n-k)!} y^{k-i-1}(1-y)^{i+n-k} dy\\
&= Pr(Beta(k-i, i+n-k+1) \leq c)
\end{aligned}
\end{equation}

Thus, via Equation (\ref{eqn:continuous}), we can  directly calculate if two mutations are closer together than by chance quickly and efficiently.

\subsection{Multiple Comparison Adjustment For Structures} \label{MultCompStruct}
In addition to the Bonferroni multiple comparison adjustment done by the NMC method, an adjustment is also required to account for testing multiple structures per protein. Since the structures for a given protein could be quite similar and thus lead to similar clustering results, a second Bonferroni adjustment would be too conservative. Instead, a combined Bonferroni-FDR approach was performed as follows. First, for a given protein, the NMC reported p-value for a given cluster was multiplied by $\frac{n(n-1)}{2}$, to calculate $P^*$. Thus, on a per-protein level, $P^*$ represents the inverse Bonferroni adjustment performed by the NMC algorithm and thus allowed us to compare each cluster's $P^*$ to an $\alpha$-level of $0.05$ to determine significance. To account for all the structures analyzed, we computed a rough FDR (rFDR) \citep{gong_atlas_2009}: $$rFDR =  \alpha * \frac{k+1}{2k}$$ where k is the total number of structures. In the case of the 1100 structures analyzed in this study, $rFDR \approx 0.02502$. Finally, any clusters for which $P* \leq 0.02502$ was deemed to be significant. For the rest of this paper, with the exception of Table \ref{tab:9Tie}, we only report the p-value to avoid confusion. Nevertheless, each cluster presented in Section  \ref{discussion} is in fact significant after adjusting for structural multiple comparisons.

\section{Results} \label{discussion}

Using the \iPAC{} package, 215 of the total 1100 structures analyzed were found to have significant clustering. When comparing \iPAC{} with the original NMC method, out of the 140 proteins analyzed, both \iPAC{} and NMC identified 8 proteins that contained significant clusters. However, \iPAC{} also identified 3 new proteins as well, specifically EGFR, EIF2AK2 and HAO1. These 3 new proteins correspond to 10 of the 215 structures found to have clustering. \iPAC{} also found structure 2ENQ for the protein PIK3CA to contain a significant cluster while NMC did not. The 8 proteins identified by both algorithms correspond to the remaining 204 structures. There were no proteins that were identified by NMC but were subsequently missed by the \iPAC{} algorithm. Please see ``Results Summary" in the supplementary materials for a full listing of which structures and which proteins were found to be significant. 

\begin{figure}[h!]
	\centering
	\includegraphics[scale=0.32]{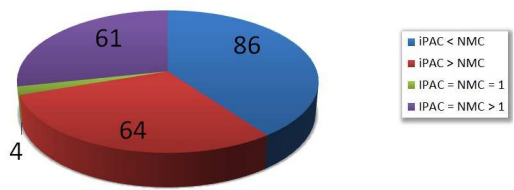}
	\caption{A comparison of NMC and \iPAC{} over all the structures that were found to be significant.}
	\label{fig:StructureTally}
\end{figure}

As can be seen from Figure \ref{fig:StructureTally}, approximately 70\% of all the structures found to have significant clustering differed in the amount of clusters identified when comparing \iPAC{} vs NMC. This leads one to believe that in some cases, consideration of the tertiary structure identifies additional clusters while in other cases, clusters are able to be removed, offering a simplified view of the mutational information.  While it is outside the scope of this paper to consider every one of the 215 structures with clustering, we present three representative cases where integration of the tertiary protein structure into the analysis had a significant effect: 1) identification of mutation clustering in a protein that would otherwise be missed, 2) identification of new mutation clusters in a protein that was detected using the NMC methodology, and 3) reduction of the total mutational clusters in a protein that was detected using the NMC methodology. We also note, as can be seen in Table \ref{tab:9Tie}, that the p-value found for the most significant cluster is similar on the protein level. Proteins that had very significant clustering, such as KRAS and TP53, remain very significant when the tertiary structure is incorporated. Proteins that were less significant, such as IDE and AKT1, remain so as well.

\begin{table}[h]
\small
    \begin{tabular}{|c|c|c|c|c|}
        \hline
        ~            &\multicolumn{2}{c|}{\iPAC{}} &  \multicolumn{2}{c|}{\emph{NMC}}   \\ \hline
 Protein &  P-value & P*  & P-value & P*\\ \hline
 KRAS   &  6.17 E-185 & 6.35 E-181 & 4.39 E-233 & 4.52 E-229\\       
 TP53   &  5.23 E-128 & 6.11 E-123 & 4.37 E-086 & 5.30 E-81\\    
 BRAF   &  3.73 E-130 & 1.01 E-126 & 3.84 E-130 & 1.04 E-126 \\       
 PIK3CA &  8.20 E-084 & 3.58 E-80  & 8.20 E-084 & 3.58 E-80\\ 
 NRAS   &  5.38 E-026 & 6.46 E-24  & 8.26 E-029 & 9.91 E-27\\ 
 HRAS   &  1.23 E-010 & 5.54 E-09  & 5.61 E-010 & 8.42 E-09 \\                        
 AKT1   &  1.18 E-005 & 7.08 E-05  & 2.47 E-005 & 7.41 E-05  \\ 
 IDE    &  2.20 E-005 & 6.60 E-05  & 1.56 E-003 & 4.67 E-03
\\ 
    \hline
    \end{tabular}
\normalsize
    \vspace{4 pt}
	\caption{A comparison of the most significant \iPAC{} and NMC p-values from the 8 proteins that were picked up by both algorithms. P* is calculated as described in Section \ref{MultCompStruct}.}
	\label{tab:9Tie}
\end{table}
\normalsize

We note that 9 out of the 11 proteins that were found significant by \iPAC{} had their most significant cluster overlap a binding site, proton acceptor site or kinase domain. For the remaining 2 proteins, the most significant cluster for PIK3CA overlapped amino acid 1047 which has been shown to ease the entrance of substrates and hence potentially increase the substrate turnover rate, a typical oncogenic behavior \citep{mankoo_pik3ca_2009}. For a detailed per protein description, please see ``Relevant Sites" in the supplementary materials. 

Finally, we considered the performance of \iPAC{} as compared to two popular machine learning algorithms, \emph{PolyPhen-2} \citep{adzhubei_method_2010} and \emph{CHASM} \citep{carter_cancer-specific_2009}. First, a direct comparison must be considered in light of the fact that these algorithms require a much more extensive set of information than \iPAC{}. Nevertheless, over 98\% of the amino acids that occurred in significant mutation clusters were also identified as significant (with a FDR of $\leq$ 20\%) by \emph{Polyphen-2} and \emph{CHASM}. For full details, please see ``Performance Comparison" in the supplementary materials.

\subsection{\iPAC{} finds novel proteins} \label{new}
\begin{figure}[h!]
	\centering
	\includegraphics[scale=0.3]{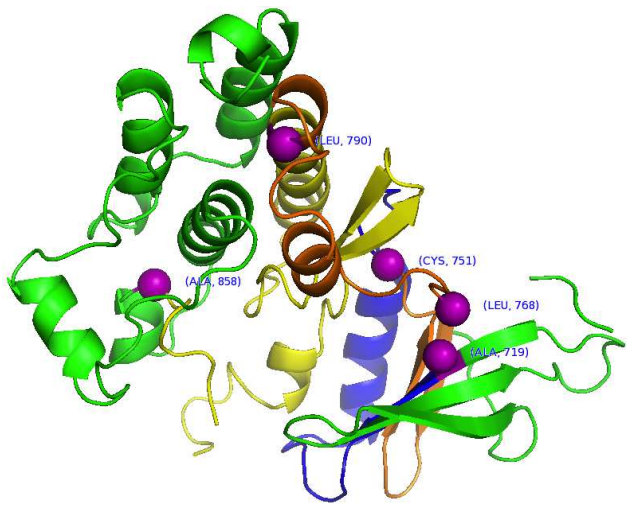}
	\caption{The 2GS7 structure color coded by region: 1) cluster 1 - orange, 2) cluster 2 - blue and 3) cluster 3 - yellow. The boundary $\alpha$-carbon amino acids of 719, 751, 768, 790 and 858 are shown as purple spheres.}
	\label{fig:EGFR-2GS7}
\end{figure}

As discussed Section \ref{discussion}, three new proteins were identified by \iPAC{} that were missed when tertiary structures are not accounted for. The EGFR protein, a cell-surface receptor for epidermal growth factor family ligands \citep{Herbst2004S21}, is perhaps the most well known and has been found in a wide array of cancers such as lung \citep{Scagliotti15062004}, anal \citep{Walker20091517} and glioblastoma multiforme \citep{Heimberger15022005}. Although seven EGFR structures were identified by \iPAC{} to contain significant clustering, we will concentrate on the 2GS7 structure \citep{Zhang20061137} as it showed the most significant clustering. As seen in Table \ref{tab:EGFRClust}, three significant clusters were found with cluster 3 being being a sub-cluster of cluster 1. Figure \ref{fig:EGFR-2GS7}, shows the orientation of these clusters in three dimensional space.

\begin{table} [h!]
\begin{center}
    \begin{tabular}{|c|c|c|c|c|}
        \hline
        Cluster  & Start & End & Muts. in Cluster & P-Value    \\ \hline
        1 & 751   & 858 & 4 & 1.35E-04 \\ 
        2 & 719   & 751 & 2 & 2.41E-03 \\ 
        3 & 790   & 858 & 2 & 2.82E-03 \\
        \hline
    \end{tabular}
\end{center}
    \caption{The three most significant clusters found in EGFR for the 2GS7 structure.}
    \label{tab:EGFRClust}
\end{table}

Overall, all the statistically significant clusters found deal with lung cancer pathology and an increase in kinase activity. The two mutations in cluster 2, G719S and T751I are both found in lung cancer with the first mutation responsible for strongly increased kinase activity \citep{ yun_structures_2007, tam_distinct_2006, paez_egfr_2004} and the second found in erlotinib responsive non small cell lung cancer patients (NSCLC) \citep{Peraldo-Neia_2011,Tsao_2005}, respectively. 
Cluster 3 contains two mutations, T790M and L858R, both of which have been found in lung cancer and are known for increased kinase activity as well \citep{yun_t790m_2008, yun_structures_2007, tam_distinct_2006, paez_egfr_2004}. Finally, cluster 1 is comprised of clusters 2 and 3, with an additional mutation S768I which potentially shows a positive clinical response to Getfinib in NSCLC patients \citep{Masago01112010}. It is interesting to note that both clusters 1 and 2, that are identified via statistical analysis, contain mutations that have been found to benefit from pharmacological intervention. Had the tertiary structure of EGFR not been taken into account, these clusters would not have been identified by the NMC algorithm. When the protein is viewed linearly, the mutations occur too far away from each other to result in statistically significant p-values.

\subsection{\iPAC{} finds additional clusters} \label{more}
\begin{figure}[h!]
	\centering
	\includegraphics[scale=0.35]{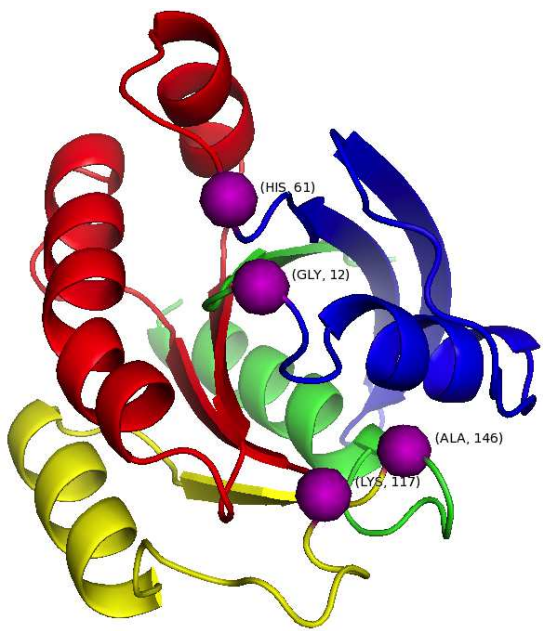}
	\caption{The 3GFT structure color coded by region: amino acids 13-60 are blue, 62-116 are red and 118-145 are yellow. The boundary $\alpha$-carbon amino acids of 12,61,117 and 146 are shown as purple  spheres. }
	\label{fig:KRAS-3GFT}
\end{figure}

One example where \iPAC{} finds additional clusters is in the KRAS protein when analyzing the 3GFT structure\footnote{For this analysis, we included included mutational and positional data only on residues 1-167. No 3D positional information was available in the 3GFT structure on residues 168-188, and these residues were removed before the analysis. Further, the structural information has amino acid 61 as a histidine (isoform 2B for KRAS in the Uniprot Database) while the COSMIC database has a glutamine in that position. However, as the substitution of one amino acid in the structure for another would not have a significant affect on its spatial orientation and as amino acid 61 has a large number of somatic mutations, it was kept in the analysis.}  \citep{kras_3GFT}. KRAS, part of the RAS set of of proteins which are involved in a large number of signaling cascades, is one of the most studied cancer oncogenes with activating mutations in approximately 17-25\% of all human cancers \citep{Kranenburg_2005}.  While both NMC and \iPAC{} identified many of the same clusters such as amino acids 12-13, 12-61 and 12-146, \iPAC{} identified several novel clusters as well, specifically amino acids 61-117 and 117 -146. We note that both algorithms specifically identify a cluster between residues 12 and 146, and given that we only have positional data for 167 residues, signifies that there is one large cluster that covers $\approx 80\%$ of all the available amino acids. However, combined with the two novel clusters identified by \iPAC{}, we are able to partition the protein into three distinct regions 1) 12-61, 2) 61 - 117 and 3) 117-146 that cover 30\%, 34\% and 18\% of the protein respectively (see Figure \ref{fig:KRAS-3GFT}).

We also ran NMC and \iPAC{} on each region separately to consider how the clustering results would be affected. As can be seen from Table \ref{tab:3GFT}, failure to account for the tertiary protein structure resulted in region 3 no longer being detected and region 1 losing significance by over ninety orders of magnitude. 

\begin{table} [h]
\begin{center}
    \begin{tabular}{|l|c|c|}
        \hline
        ~       & \multicolumn{2}{c|}{P-value}      \\ \hline
        Region  & NMC           & \iPAC{}           \\ 
        1) 12-61      & 1.37E-11 & 3.36E-105 \\ 
        2) 61-117  & -             & -              \\ 
        3) 117-146 & -             & 3.35E-12 \\
        2\&3) 61 - 146 & - & 3.31E-05 \\
        \hline
    \end{tabular}
    \end{center}
    \caption{P-value for each region when the region is considered independently. A ``-" signifies that the region was not found to be significant.}
	\label{tab:3GFT}
\end{table}

Further, while somatic mutations in region 12-61 have been found in many cancers such as colorectal, lung, pancreatic and bladder \citep{sjoblom_consensus_2006,  tam_distinct_2006,lee_clinicopathologic_1995,motojima_detection_1993,nakano_isolation_1984, santos_malignant_1984}, somatic mutations at amino acids 61, 117 and 146 have primarily been found in lung and colorectal carcinomas. Even more specifically, mutations at amino acids 117 and 146 (K $\rightarrow$ N and A $\rightarrow$ T, respectively) deal mostly with colorectal cancer \citep{sjoblom_consensus_2006}. Thus, by taking into account the tertiary structure, the clusters identified by \iPAC{} subdivide the protein along pathological lines. 

\subsection{\iPAC{} finds fewer clusters than NMC} \label{less}
Of the 215 structures found to contain significant clustering, 86 structures were identified where \iPAC{} found fewer structures than NMC. Three of these structures correspond to BRAF, 31 correspond to HRAS and 52 correspond to TP53. Here, we consider structure 3TV4 \citep{wenglowsky_2011} for the BRAF protein as it contains the most significant cluster found by both \iPAC{} and NMC. For this protein, it is well known that amino acid 600 is one of the most highly mutated residues. In our dataset, 60 of the 76 total mutations that fulfilled the requirements described in section \ref{data:COSMIC} occurred on amino acid 600. As expected, the most significant ``cluster" is located solely on that amino acid, with an \iPAC{} p-value of $3.73\times10^{-130}$ and an NMC p-value of $3.84\times10^{-130}$. However, in total, \iPAC{} identifies 9 clusters for this structure while NMC identifies 19, with the differences shown in Table \ref{tab:BRAF}.

\begin{table}[htb!]
\centering
\subfloat[Clusters found by both NMC and \iPAC{}]{
    \begin{tabular}{|c|c|c|c|c|}
        \hline
        ~     & ~   & ~                & \multicolumn{2}{c|}{P-value}                      \\ 
        Start & End & \# Muts. & iPAC                    & NMC                     \\ \hline       
        600   & 600 & 60               & 3.73 E-130 & 3.84 E-130 \\ 
        469   & 600 & 70               & 9.76 E-122 & 5.63 E-16  \\ 
        600   & 601 & 62               & 3.10 E-79  & 1.35 E-117 \\ 
        597   & 600 & 62               & 4.05 E-77  & 2.20 E-105 \\ 
        464   & 600 & 71               & 1.25 E-73  & 1.74 E-16  \\ 
        596   & 600 & 64               & 3.06 E-73  & 8.28 E-103 \\ 
        581   & 600 & 66               & 1.99 E-51  & 2.96 E-64  \\ 
        600   & 671 & 63               & 7.78 E-15  & 3.54 E-28  \\ 
        469   & 469 & 4                & 7.50 E-04  & 7.50 E-04   \\
        \hline
    \end{tabular}
    \label{tab:BRAFboth}
}
    
\qquad
\subfloat[Clusters dropped by \iPAC{}]{
    \begin{tabular}{|c|c|c|c|}
        \hline
        Start & End & \# Muts. & NMC Pvalue              \\ \hline
        597   & 601 & 64      & 8.28 E-103  \\ 
        596   & 601 & 66      & 9.97 E-102 \\ 
        581   & 601 & 68      & 8.73 E-67  \\ 
        596   & 671 & 67      & 1.10 E-31  \\ 
        597   & 671 & 65      & 1.93 E-29  \\ 
        581   & 671 & 69      & 2.22 E-28  \\ 
        464   & 601 & 73      & 7.09 E-19  \\ 
        469   & 601 & 72      & 3.58 E-18  \\ 
        464   & 671 & 74      & 6.01 E-09  \\
        469	  & 671 & 73      & 2.38 E-08 \\
        \hline
    \end{tabular}
	\label{tab:BRAFNMC}
}

\qquad
\caption{The significant clusters found by both \iPAC{} and NMC are shown in Table \ref{tab:BRAFboth}. The clusters that were not deemed significant by \iPAC{} but were deemed significant by NMC are shown in Table \ref{tab:BRAFNMC}.}
\label{tab:BRAF}
\end{table}

\begin{figure}[htb!]
	\centering
	\includegraphics[scale=0.30]{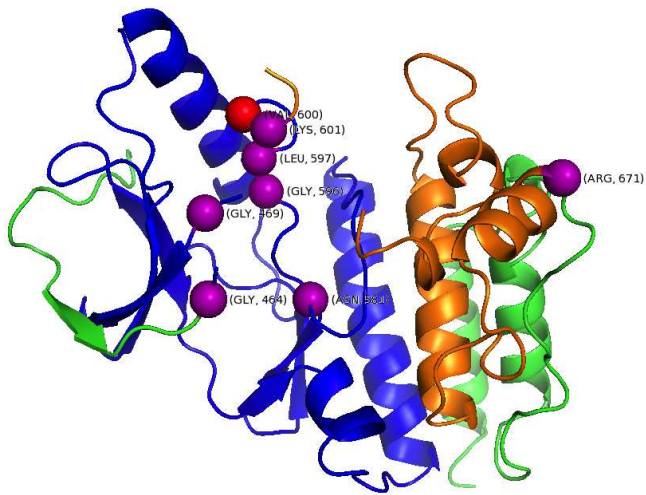}
	\caption{The 3TV4 structure color coded by region: 1) Amino 464-600 are blue 2) Amino Acids 601-671 are orange. The  $\alpha$-carbons of the mutated amino acids 464, 466, 469, 581, 596, 597, 601 and 671 are shown as purple spheres. Amino acid 600 is colored red.}
	\label{fig:BRAF-3TV4}
\end{figure}

While it is outside the scope of this paper to consider all the differences between Tables \ref{tab:BRAFboth} and \ref{tab:BRAFNMC}, we would like to point out that, contrary to \iPAC{}, the NMC algorithm reports the two longest clusters: 1) 464 - 671 (p-value = $6.01 \times 10^{-9}$ ) and 2) 469-671 (p-value = $2.38 \times 10^{-8}$). After alignment of the structure as described in Section \ref{data:PDB}, we only have structural information on amino acids 448 - 723. Thus, the largest cluster detected by NMC covers $\approx 75\%$ of all the amino acids that we are considering. However, by taking into account the 3D structure of the protein, these ultra-long clusters are dropped and the clusters where \iPAC{} and NMC overlap show 2 distinct areas of the protein, amino acids 464-600 and 600-671. As expected, as the majority of mutations occur on amino acid 600, both NMC and \iPAC{} declare that the ``cluster" located at amino acid 600 is highly significant.

Further, as described below, by considering only the clusters when taking into account the 3D structure (see Figure \ref{fig:BRAF-3TV4}), the results again tend to fall along pathological function. After applying the methodology described in Section \ref{data:COSMIC}, the mutations that were found to be in significant clusters included G464V, G466V, G469V, G469A, N581S, G596R, L597V, LV597R, V600E, V600K, K601N and R671Q. As R671Q was found in only one sample within the COSMIC database and does not have extensive literature, we will not include it in further discussion. Taking into account the 3 most significant clusters picked up by \iPAC{} and NMC, we now consider the protein in 3 parts: A) Residues 469 - 599, B) Residue 600 and C) Residue 601 (we have slightly adjusted the clusters displayed in Table \ref{tab:BRAFboth} to avoid overlap). The mutations listed that fall with region A, correspond primarily to lung and colorectal cancer \citep{Jeet_2009,pratilas_genetic_2008,greenman_patterns_2007, lee_braf_2003, davies_mutations_2002,naoki_missense_2002}. Region B, which is comprised of only amino acid 600 is by far the most common mutation with BRAF. This mutation results in constitutive and elevated kinase activity and has been found in a large range of cancers including colorectal carcinoma, ovarian serous carcinoma, metastatic melanoma and pilocytic  astrocytoma. Further, supporting the hypothesis that somatic clusters might provide pharmacological targets, it has already been shown that suppression of this cluster in melanoma causes tumor growth arrest and helps promote apoptosis \citep{andreu-perez_protein_2011,	greenman_patterns_2007,	sjoblom_consensus_2006,hingorani_2003, rajagopalan_2002, davies_mutations_2002}. Finally, the K601N mutation in region C has been found in multiple myeloma patients who also may benefit from BRAF inhibitors \citep{chapman_initial_2011}.  

\section{Conclusion}

In this paper, we extended the existing methodology available to find somatic mutation clustering by utilizing the information provided in the protein tertiary structure. In doing so, we showed that we are able to find both new proteins with clustering as well as new clusters in previously found proteins. We have also shown that by taking into account 3D structure, we are able to remove clusters that do not have biological meaning. The method is fast and robust, with the vast majority of proteins analyzed within 5-10 minutes when executed on a desktop with 8 GB of DDR3 RAM and an Intel i7 3600k processor running at a frequency of 3.40 GHZ. Further, as the underlying calculation relies upon the NMC algorithm, a preset fixed window size is not required which allows for the detection of clusters of various lengths \citep{ye_2010}. We have also shown that by employing a completely statistical methodology, we are able to identify mutations that when may be suppressed via pharmacological intervention and stop further tumor growth.

This methodology, while an improvement on the NMC method, still suffers from some limitations. First, the mutation status of all the amino acids must be determined although with the advent of high-throughput sequencing, this will become less of an issue as time progresses. Also, both hypermutability of genomic locations and unequal rates of mutagenesis might violate the assumption that each amino acid has a uniform mutation probability. For instance, it is well known that hypermutable positions for both somatic and germline mutations exist.  Insertions and deletions that are typically sequence dependent have been removed from the analysis and only missense substitutions of single amino acids have been kept in this study to help reduce such uniformity violations. Similarly, CpG dinucleotides can have mutational frequency that is ten times or more that of other dinucleotides \citep{Sved01061990}. However, less than 13\% of the mutations used to find clustering in Sections \ref{new}, \ref{more}, and \ref{less} were in CpG sites. Further, as described by \citet{ye_2010}, tobacco smoking preferentially causes transversions in lung cancer while the mutational landscape for colorectal cancer has more transitions \citep{hollstein_p53_1991}. Nevertheless, in the context of KRAS, the vast majority of mutations occur on amino acids 12, 13 and 61 for both lung and colorectal cancer. This suggests that while the mutational spectrum may be different, it does not have a large effect on the position of mutations and thus the uniformity assumption. As with previous studies, while this analysis is influenced by nonrandom factors, it nonetheless appears that selection of a cancer phenotype is the primary cause of clustering. 

It should also be noted that while \iPAC{} is designed to take tertiary structure into account, it is only able to do so by appealing to the MDS methodology. Future research is required in order to relax this restriction to potentially identify additional clustering results. Finally, as shown in Section \ref{discussion}, \iPAC{} finds fewer clusters for a significant percentage of the structures analyzed. This reduction in total clusters can come from two sources: the removal of some amino acids due to lack of tertiary position information or that the cluster is no longer found to be significant when 3D structure is taken into account. The first source, while already rare will become even more so in the future as more detailed structural information becomes available. As for the second source, when a cluster is not identified under \iPAC{} when compared to \emph{NMC}, an overlapping or nearby cluster is typically found (as shown in Tables \ref{tab:BRAFboth} and \ref{tab:BRAFNMC}). For BRAF specifically, there was a total of 3 structures where \iPAC{} found fewer clusters than NMC. Further,  every ``possibly" or ``probably damaging" mutation, as categorized by PolyPhen-2 \citep{adzhubei_method_2010}, was still represented in at least one cluster in each structure. Thus, in the case of BRAF, none of the damaging mutations identified by PolyPhen-2 were lost. For a more detailed analysis, please see ``Potential Driver Loss" in the supplementary materials. Ultimately, further research is required to further reduce the possibility of losing driver mutations while taking into account tertiary structure.

In conclusion, we present a novel approach to identifying mutation clustering while taking into account protein tertiary structure. We further show that by taking into account tertiary structure we are able to detect clusters that would otherwise be missed. Next, we demonstrate that for some of the clusters found, pharmacological intervention has already been successfully applied, further confirming the hypothesis that mutational clustering might point to activating driver mutations. As additional protein structures continue to be solved, \iPAC{} would be able to rapidly perform a statistical analysis to identify such potential mutations. Finally, as we gain a better understanding of the tertiary structure of DNA, this method might also have applications  to finding mutational clustering on the DNA level.

\section*{Acknowledgements}
We thank Drs. Francesca Chairomonte and Catherine Siena Grasso for their time and discussions regarding the development of this methodology.

\section*{Funding Sources:}
This work was supported in part by NSF Grant DMS 1106738 (G. Ryslik, H. Zhao), NIH Grant GM59507 (H. Zhao), and the China Scholarship Council (Y. Cheng).

\bibliography{refs}{}
\bibliographystyle{natbib}
\end{document}